%Paper: hep-ph/9305260
%From: luke@yukawa.UCSD.EDU
%Date: Wed, 12 May 93 16:25:35 PDT

\input epsf

\ifx\epsffile\undefined\message{(FIGURES WILL BE IGNORED)}
\def\insertplot#1#2{}% null macro
\def\insertfig#1#2{}% null macro
\else\message{(FIGURES WILL BE INCLUDED)}
\def\insertplot#1#2{
\midinsert\centerline{{#1}}\vskip0.2truein\centerline{{\epsfxsize=\hsize
\epsffile{#2}}}\vskip0.5truecm\endinsert}
\def\insertfig#1#2{
\midinsert\centerline{{\epsfxsize=\hsize\epsffile{#2}}}\vskip0.2truein
\centerline{{#1}}\vskip0.5truecm\endinsert}
\fi

\input harvmac
%%%%%%%%%%%%%%%%%%%%%%%%%%%%%%%%%%%%%%%%%%%%%%%%%%%%%%%%%%%%%%%%%%%%%%
%
%  UCSD macros to overwrite some of the definitions in harvmac.tex
%  (include after harvmac.tex)
%  last modified 4/92
%
%%%%%%%%%%%%%%%%%%%%%%%%%%%%%%%%%%%%%%%%%%%%%%%%%%%%%%%%%%%%%%%%%%%%%%%
%
% modify the output routine for the little format
%
\ifx\answ\bigans
\else
\output={
  \almostshipout{\leftline{\vbox{\pagebody\makefootline}}}\advancepageno
}
\fi
%
%
% address
%

%
% grant numbers
%

%
% preprint number
%
\def\UCSD#1#2{\noindent#1\hfill #2%
\bigskip\supereject\global\hsize=\hsbody%
\footline={\hss\tenrm\folio\hss}}% restores pagenumbers
%
% abstract
%
\def\abstract#1{\centerline{\bf Abstract}\nobreak\medskip\nobreak\par #1}
%
%
% titlefont
%
%
\edef\tfontsize{ scaled\magstep3}
 \tfontsize  \tfontsize
 \tfontsize \font\titlei=cmmi10 \tfontsize
\font\titleis=cmmi7 \tfontsize \font\titleiss=cmmi5 \tfontsize
\font\titlesy=cmsy10 \tfontsize \font\titlesys=cmsy7 \tfontsize
\font\titlesyss=cmsy5 \tfontsize  \tfontsize
\skewchar\titlei='177 \skewchar\titleis='177 \skewchar\titleiss='177
\skewchar\titlesy='60 \skewchar\titlesys='60 \skewchar\titlesyss='60
%
%\def\titlefont{\def\rm{\fam0\titlerm}% switch to title font
%\textfont0=\titlerm \scriptfont0=\titlerms \scriptscriptfont0=\titlermss
%\textfont1=\titlei \scriptfont1=\titleis \scriptscriptfont1=\titleiss
%\textfont2=\titlesy \scriptfont2=\titlesys \scriptscriptfont2=\titlesyss
%\textfont\itfam=\titleit \def\it{\fam\itfam\titleit}\rm}
%
%
% math symbols
%
%---------------------------------------------------------------------
%
\def\inv{^{\raise.15ex\hbox{${\scriptscriptstyle -}$}\kern-.05em 1}}
  %prime
\def\lbar{{\lower.35ex\hbox{$\mathchar'26$}\mkern-10mu\lambda}} %lambda bar

%
%
% various slashed symbols
%
%
\def\slash#1{\rlap{$#1$}/} % slashes a character
\def\dsl{\,\raise.15ex\hbox{/}\mkern-13.5mu D} %this one can be subscripted
\def\delsl{\raise.15ex\hbox{/}\kern-.57em\partial}
\def\Ksl{\hbox{/\kern-.6000em\rm K}}
\def\Asl{\hbox{/\kern-.6500em \rm A}}
\def\Dsl{\hbox{/\kern-.6000em\rm D}} %roman D
\def\Qsl{\hbox{/\kern-.6000em\rm Q}}
\def\gradsl{\hbox{/\kern-.6500em$\nabla$}}
%
% space and backspace in l mode
%
\def\lspace{\ifx\answ\bigans{}\else\qquad\fi}
\def\lbspace{\ifx\answ\bigans{}\else\hskip-.2in\fi} % $$\lbspace...$$
%
%     boxes an equation
%
\def\boxeqn#1{\vcenter{\vbox{\hrule\hbox{\vrule\kern3pt\vbox{\kern3pt
        \hbox{${\displaystyle #1}$}\kern3pt}\kern3pt\vrule}\hrule}}}
%
%     draw a little box (end of proof symbol)
%     e.g. \mbox{.1}{.1}
%
\def\mbox#1#2{\vcenter{\hrule \hbox{\vrule height#2in
\kern#1in \vrule} \hrule}}
%
%
%
%     curly letters
%
   %curly letters

%
%
%
%     derivatives
%
%

%

\def\bar#1{\overline{#1}}

\def\darr#1{\raise1.5ex\hbox{$\leftrightarrow$}\mkern-16.5mu #1}

%
 %pound sterling
%
 %puts a small half in a displayed eqn
\def\frac#1#2{{\textstyle{#1\over #2}}} %puts a small fraction
%in a displayed eqn
%
%
%     various math operators
%
%
\def\tr{\mathop{\rm tr}}

\def\MeV{{\rm MeV}}

%
%
%
%

%
%       relations
%
\def\ltap{\ \raise.3ex\hbox{$<$\kern-.75em\lower1ex\hbox{$\sim$}}\ }
\def\gtap{\ \raise.3ex\hbox{$>$\kern-.75em\lower1ex\hbox{$\sim$}}\ }
\def\gl{\ \raise.5ex\hbox{$>$}\kern-.8em\lower.5ex\hbox{$<$}\ }
\def\roughly#1{\raise.3ex\hbox{$#1$\kern-.75em\lower1ex\hbox{$\sim$}}}
%
%
%       This defines et al., i.e., e.g., cf., etc.

\def\etal{\hbox{\it et al.}}

\def\np#1#2#3{{Nucl. Phys. } B{#1} (#2) #3}

\def\physrev#1#2#3{{Phys. Rev. } {#1} (#2) #3}
\def\ap#1#2#3{{Ann. Phys. } {#1} (#2) #3}

\relax

\def\({\left(}
\def\){\right)}
\def\cbar{\bar c}

\noblackbox
%\draftmode
 at 12truept
\vskip 1.in
\centerline{{\titlefont{Heavy Quark Fragmentation to}}}
\medskip
\centerline{{\titlefont{Polarised Quarkonium}}}
\bigskip
\centerline{Adam F.~Falk{}$^{a}$
\footnote{$^\dagger$}{Address after
10/1/93: Dept. of Physics, UC San Diego, La Jolla, CA 92093},
Michael Luke{}$^{b}$
\footnote{$^\ddagger$}{Address after
9/1/93: Dept. of Physics, University of Toronto, Toronto, Ontario,
Canada M5S 1A7},
Martin J.\ Savage{}$^{b}$
\footnote{$^{\S}$}{Address after 9/1/93: Dept. of Physics,
Carnegie Mellon University, Pittsburgh, PA 15213}
\footnote{$^{\star}$} {SSC Fellow}
and Mark B. Wise{}$^{c}$}
\bigskip
\centerline{\sl a) Stanford Linear Accelerator Center, Stanford CA 94309}
\centerline{\sl b) Department of Physics, University of California at San
Diego,}\centerline{\sl 9500 Gilman Drive, La Jolla CA 92093}
\centerline{\sl c) California Institute of Technology, Pasadena CA 91125}
\vfill
\abstract{We calculate the polarisation of $\psi$'s and $\Upsilon$'s
produced by
the fragmentation of heavy quarks.  We find that fragmentation to
transversely aligned quarkonium is slightly enhanced relative to
longitudinally polarised.  This net alignment corresponds to a
$\sim 5\%$
asymmetry in the angular distribution of leptons produced in the
subsequent decay of the quarkonium.  We point out that the leading
gluon contribution to $\psi$ and $\Upsilon$ production
arises from the evolution of the charm quark fragmentation function,
rather than from direct gluon fragmentation.}
\vfill
\UCSD{\vbox{
\hbox{UCSD/PTH 93-06, CALT-68-1864, SLAC-PUB-6175}
\hbox{hep-ph/9305260}}}{May 1993}
\eject

Quarkonium production in high energy processes is dominated by the
fragmentation of heavy quarks and gluons
\ref\kuhn{J.H. K\"uhn and H. Schneider, \physrev{D24}{1981}{2996}}%
\nref\barga{
V. Barger, K. Cheung and W.Y. Keung, \physrev{D41}{1990}{1541}}%
\nref\erica{E. Braaten and T.C. Yuan, NUHEP-TH-92-23,
UCD-92-25 (1992)}--\ref\ericb {E. Braaten, K. Cheung and T.C. Yuan,
NUHEP-TH-93-2,UCD-93-1 (1993)}.  For example, in $Z^0$ decay the short
distance process $Z^0\rightarrow\psi g g$ is suppressed relative to the
fragmentation process $Z^0\rightarrow\psi c\cbar$ by a factor of
$m_c^2/m_Z^2$, corresponding to the small overlap of the $\psi$
wavefunction with a $c\cbar$ pair that has relative momentum of order
$m_Z$.

Recently it has been shown by Braaten and Yuan \erica\ and Braaten,
Cheung and Yuan \ericb\ that the process independent fragmentation
functions for quarkonium production in high energy processes
are calculable.
For example, in $Z^0$ decay to a $\psi$ of energy $E\gg
m_\psi$, the differential decay rate may be written in the factorised
form
\eqn\factorise{d\Gamma(Z^0 \rightarrow \psi(E) + X)
=\sum_i \int_0^1 dz \, d{\widehat \Gamma}(Z^0 \rightarrow i(E/z) +
X,\mu)
\; D_{i \rightarrow \psi}(z,\mu).}
Here $d{\widehat \Gamma}(Z^0 \rightarrow i(E/z) + X,\mu)$ is the
differential decay rate for a $Z^0$ to decay to an on-shell parton of
type $i$, while the fragmentation function $D_{i \rightarrow
\psi}(z,\mu)$ gives the probability for the parton to split into a $\psi$
with momentum fraction $z$ and another parton with momentum fraction
$1-z$.  The advantage of this factorised form is that the fragmentation
function is independent of the production process.  QCD
corrections of the form $\log(m_Z/m_c)$
necessitate the introduction of a factorisation scale
$\mu$; large logarithms of $\mu/m_c$ in $D_{i \rightarrow \psi}(z,\mu)$
may be summed using the Altarelli-Parisi equations \ref
\ap{G.~Altarelli and G.~Parisi, \np{126}{1977}{298}}\ref\field
{R.~B.~Field, Applications of Perturbative
QCD, Addison Wesley (1989)}.

The fragmentation functions
$D_{c \rightarrow \psi}(z,\mu)$ and
$D_{b \rightarrow \Upsilon}(z,\mu)$ were calculated
in \ericb, explicitly
summing over quarkonium polarisations.
However, the fragmentation functions for transverse and
longitudinally polarised quarkonium are individually calculable and are
interesting in their own right.
%First of all, they are a
%testable prediction of perturbative QCD.  Second, since fragmentation
%to $\psi$'s is a potential
%background for the rare decay
%$b\rightarrow \psi s$, a theoretical understanding of the
%$\psi$ polarisation will be useful in distinguishing the two processes.
In this work we compute the fragmentation function
for production of transversely polarised $\psi$'s from $c$
quarks and $\Upsilon$'s from $b$ quarks.

The Feynman diagrams responsible for the fragmentation of
a heavy quark into quarkonium are shown in
\fig\feyn{Feynman diagrams responsible for the fragmentation
$c\rightarrow\psi$.}, where the black circle
represents some quark-antiquark
production process. For definiteness we consider
$Z^0$ decay to $\psi c\cbar$; however
the fragmentation function we will derive is independent of the
production process and is trivially extended to $b$ fragmentation
to $\Upsilon$.
Factorisation is only manifest in $\bar q\cdot
A=0$ gauge (where $\bar q$ is the antiquark four-momentum),
in which the diagram in \feyn(b) vanishes to leading order
in $m_c/E$.  Following
the manipulations in Eqs.~(9)-(11) of Ref.~\ericb,
we write the decay rate
$\Gamma(Z^0\rightarrow\psi(\epsilon) c\cbar)$, averaging over
initial spins and summing over colours, as
\eqn\totalrate{\eqalign{\Gamma({Z^0\rightarrow\psi(\epsilon)
c\cbar})=
&{1\over 2 m_Z}\int\,[dq]\,[d{\bar q}]\,(2\pi)^4\delta^{(4)}
(p_Z-{\bar q}-q)\cr &\qquad\times\int_0^1\,{dz\over 8\pi}
\int_0^\infty
\,{ds\over 2\pi}\,\theta\(s-{4m_c^2\over z}-{m_c^2\over 1-z}\)
\vert A(\epsilon) \vert^2,}}
where $A(\epsilon)$ is the amplitude to
produce a $\psi$ with polarisation $\epsilon$,
$p_Z$ is the $Z^0$ four-momentum,
$s=q^2$ is the invariant mass of the original $c$ quark,
$[dq]\equiv d^3q/(16\pi^3 q_0)$, and $z$ is the
longitudinal momentum fraction of the $\psi$.
Choosing a frame such that
$q=(q^0,0,0,q^3)$, we have
$z=(p_0+p_3)/(q_0+q_3)$, where $p$ is the $\psi$ four-momentum.

Squaring the result from \feyn\ and using the identity
\eqn\spinsum{\sum\limits_T \epsilon^T_i
\epsilon^{*T}_j=\delta_{ij}-{p_i p_j\over \vec{p}{}^2}}
to sum over transverse $\psi$ polarisations, we find
\eqn\transverse{\sum\limits_T\vert A(\epsilon)\vert^2=
{1\over(s-m_c^2)^4}D^T(s,z)
\(-g^{\alpha\beta}+{p_Z^\alpha p_Z^\beta\over m_Z^2}\)
\tr\(\Gamma_\alpha\slash{\bar q}\Gamma_\beta \slash{q}\),}
where
\eqn\dformula{\eqalign{D^T(s,z)= {256\over 81}\pi^2
\alpha_s^2 f_\psi^2
&\left[8m_c^2 s\(z-2+{2\over z}\)-
8 m_c^4\(z+4-{6\over z}+{8\over z^2}\)\right.\cr
&\qquad\left.
+{16m_c^2(s-m_c^2)z^2\over 2-z}+{8(s-m_c^2)^2z^2(1-z)\over (2-z)^2}
\right].}}
We have dropped terms suppressed by $m_c/E$ in our result.
$f_\psi$ is the $\psi$ decay constant, defined by
\eqn\fpsi{\langle 0\vert \cbar\gamma^\mu c\vert\psi(p,\epsilon)
\rangle=f_\psi m_\psi\epsilon^\mu.}
This is related to the nonrelativistic radial wavefunction at the
origin, $R_\psi(0)$,
by $f_\psi=\sqrt{3/\pi m_\psi}\,R_\psi(0)$.  Numerically,
from the $\psi$ leptonic width, $f_\psi\approx 410\,\MeV$.
The total decay rate to transversely aligned $\psi$'s may now be
written in
the factorised form
\eqn\factorrate{\eqalign{\Gamma(Z^0\rightarrow\psi^T c\cbar)=
&{1\over 16\pi^2}\Gamma(Z^0\rightarrow c\cbar)\cr
&\quad\times \int_0^1dz
\int_0^\infty ds
\,\theta\(s-{4m_c^2\over z}-{m_c^2\over 1-z}\)
{1\over(s-m_c^2)^4} D^T(s,z),}}
where the decay rate to free quarks is
\eqn\freequarks{\eqalign{\Gamma(Z^0\rightarrow c\cbar)=&
{1\over 2m_Z}\int\,[dq]\,[d\bar q](2\pi)^4\delta^{(4)}(p_Z-q-\bar q)\cr
&\qquad\times\(-g^{\alpha\beta}+{p_Z^\alpha p_Z^\beta\over m_Z^2}\)
\tr\(\Gamma_\alpha \slash{\bar q}\Gamma_\beta \slash{q}\).}}
Performing the integral over $s$
in \factorrate, we find the expression for the
fragmentation function to transversely aligned $\psi$'s:
\eqn\tranb{\eqalign{
D^T_{c\rightarrow\psi}(z,\mu=3m_c) = &{16\over 81 m_c^2}\alpha_s(3m_c)^2
f_\psi^2\cr
&\qquad\times{2\over 3}{z(1-z)^2\over (2-z)^6}
\left( 16-32z+76z^2-36z^3+6z^4\right).}}
Adding the longitudinal polarisation, we recover the
total fragmentation function $D^{T+L}_{c\rightarrow\psi}(z)$
given in \ericb,
\eqn\tot{\eqalign{
D^{T+L}_{c\rightarrow\psi}(z,\mu=3m_c) = &{16\over 81 m_c^2}
\alpha_s(3m_c)^2 f_\psi^2\cr
&\qquad\times{z(1-z)^2\over
(2-z)^6}\left(16-32z+72z^2-32z^3+5z^4\right).}}

In experiments performed at a hadron collider, the gluon fragmentation
function also plays an important role.  The Altarelli-Parisi evolution
of $D^{T\,(T+L)}_{c\rightarrow\psi}(z,\mu)$ from the scale $\mu=3m_c$
to a high energy scale $\mu=M$ (such as $M=m_Z$) will induce a gluon
fragmentation function $D^{T\,(T+L)}_{g\rightarrow\psi}(z,M)$ via the
gluon splitting function $P_{g\to c\bar c}(z)$.  That is, a high energy
gluon may split into a quark-antiquark pair of lower energy, either of
which may then fragment to quarkonium.  In the leading logarithmic
approximation this effect, of order $\alpha_s^3\ln(M/m_c)$, dominates
over the direct contribution to gluon fragmentation computed in \erica,
which is of order $\alpha_s^3$ without the $\ln(M/m_c)$
enhancement.\footnote{$^\dagger$}{This applies only to $\psi$ and
$\Upsilon$ production; gluon fragmentation to $\eta_c$'s and $\eta_b$'s
is of order $\alpha_s^2$, and so the direct gluon fragmentation
calculated in \erica\ dominates.}

It is equally straightforward to evolve the quark fragmentation functions
\tranb\ and \tot\ to high energies using the Altarelli-Parisi equations
and the quark splitting functions $P_{c\to cg}(z)$ and $P_{c\to gc}(z)=
P_{c\to cg}(1-z)$.  Because $\int_0^1 dz\ P_{c\to cg}(z)=0$, this
evolution softens the $z$ distributions of
$D^{T}_{c\rightarrow\psi}(z,\mu)$ and
$D^{T+L}_{c\rightarrow\psi}(z,\mu)$ but leaves their integrals
$\int_0^1 dz\ D^{T\,(T+L)}_{c\rightarrow\psi}(z,\mu)$ unchanged \ericb.

We now define $\zeta$
to be the ratio of transverse to total fragmentation probabilities,
\eqn\zet{\zeta = {\int_0^1 dz\ D^T_{c\rightarrow\psi}(z,\mu)
\over \int_0^1
dz\ D^{T+L}_{c\rightarrow\psi}(z,\mu)}.}
As discussed in the previous paragraph, $\zeta$ is independent of
$\mu$.
Evaluating \zet\ at $\mu=3m_c$,
we find $\zeta=0.69$, to be compared with $\zeta=\frac23$ for
production of unaligned $\psi$'s.
Therefore a small excess of transversely aligned $\psi$'s will be
produced.  Since the
dependence on $f_\psi$ and $m_c$ drops out of \zet, the ratio $\zeta$
is the same for $\psi$'s and $\Upsilon$'s.
In leading logarithmic approximation, the
corresponding ratio for gluon fragmentation
to $\psi$'s has the same value and
is $\mu$ independent, since the gluon fragmentation functions
are induced only by quark fragmentation.
Hence, at a hadron collider, where $\psi$'s are produced both
by quark and gluon fragmentation, the fraction of transversely
aligned $\psi$'s is also $\zeta=0.69$.

The asymmetry $\zeta$ is measurable through the angular distribution
of the leptons in the decay
$\psi\rightarrow\ell^+\ell^-$.  Defining $\theta$ to
be the angle between the alignment axis and the lepton momentum, the
angular distribution $d\Gamma/d\theta$ in the $\psi$ rest
frame has the form
\eqn\defalph{{d\Gamma(\psi\rightarrow\ell^+\ell^-)\over d\theta}
\propto 1+ \alpha\cos^2\theta}
where
\eqn\givealph{\alpha={3\zeta-2\over 2-\zeta}.}
$\zeta=0.69$ corresponds to $\alpha=0.053$, about a $5\%$ asymmetry.

It is useful to compare this with the polarisation of $\psi$'s
produced from nonleptonic $b$ decay.  Assuming the four-quark
amplitude factorises
(which should give a reasonable estimate, as the $\psi$ is a compact
object on the hadronic scale $\Lambda_{QCD}$), a
straightforward calculation of the decay
$b\rightarrow\psi s\rightarrow\ell^+\ell^- s$ gives
\eqn\bdec{\alpha=-{m_b^2-m_\psi^2\over m_b^2+m_\psi^2}\approx -0.46.}
This is in nice agreement with a recent CLEO
inclusive measurement, which finds
\ref\cleomeas{The CLEO Collaboration (A.~Bean \etal), ``Decay
Rates and Polarization in Inclusive and Exclusive $B\rightarrow\psi$
Decays'', Talk presented at the {\sl 1992 International Conference on
High Energy Physics}, Dallas, Texas} (see
also \ref\argus{H. Schr\"oder, ``ARGUS Results on
$B$ Decays via $b\rightarrow c$ Transitions'', published
in
{\it Proceedings of the 25th International Conference
on High Energy Physics}, K.~K.~Phua and Y.~Yamaguchi, eds.,
World Scientific (1991)})
\eqn\cleo{\alpha=-0.44\pm 0.16}
(where we have added the systematic and statistical
errors linearly).  Since the inclusive branching ratio for
$b\rightarrow\psi X$
is $(1.02\pm0.05\pm0.09)\%$ \cleomeas, this decay may
provide a significant background to $\psi$ production by fragmentation
at high energy colliders.
There is no comparable decay process to compete with $b$ quark
fragmentation
for $\Upsilon$ production; therefore this mode should allow a cleaner
measurement of $\zeta$.

We note that
$\psi$ (or $\Upsilon$) production can also arise from the fragmentation
into P-wave quarkonium which subsequently decays electromagnetically
to $\psi$  (or $\Upsilon$), as mentioned in \ericb.
For weakly bound nonrelativistic quarkonium, the fragmentation into
P-wave states is suppressed; however, in the charmonium system
this may be an important additional source of $\psi$'s.

In summary, we have calculated the  leading contribution to the
transverse fragmentation function for
quarkonium production from heavy quarks.  We find that the
$\psi$ and $\Upsilon$ are essentially unaligned, with only a $\sim 5\%$
angular anisotropy in the lepton distribution produced in their
electromagnetic decays.
We also point out that the dominant contribution to gluon fragmentation
into $\psi$ or $\Upsilon$  in high energy processes
arises from Altarelli-Parisi evolution of heavy quark
fragmentation functions, rather than from direct gluon fragmentation.

\bigskip

This work was supported in part by the Department of Energy under contracts
DE--AC03--76SF00515 (SLAC), DE--FG03--90ER40546 (UC San Diego) and
DE--FG03--92ER40701 (Caltech).
MJS acknowledges the support of a Superconducting Supercollider
National Fellowship from the Texas National Research Laboratory
Commission under grant  FCFY9219.

\vfil\eject

\listrefs
%\listfigs
\vfil\eject
\insertfig{Fig.~1.  Feynman diagrams responsible for the fragmentation
$c\to\psi c$.}{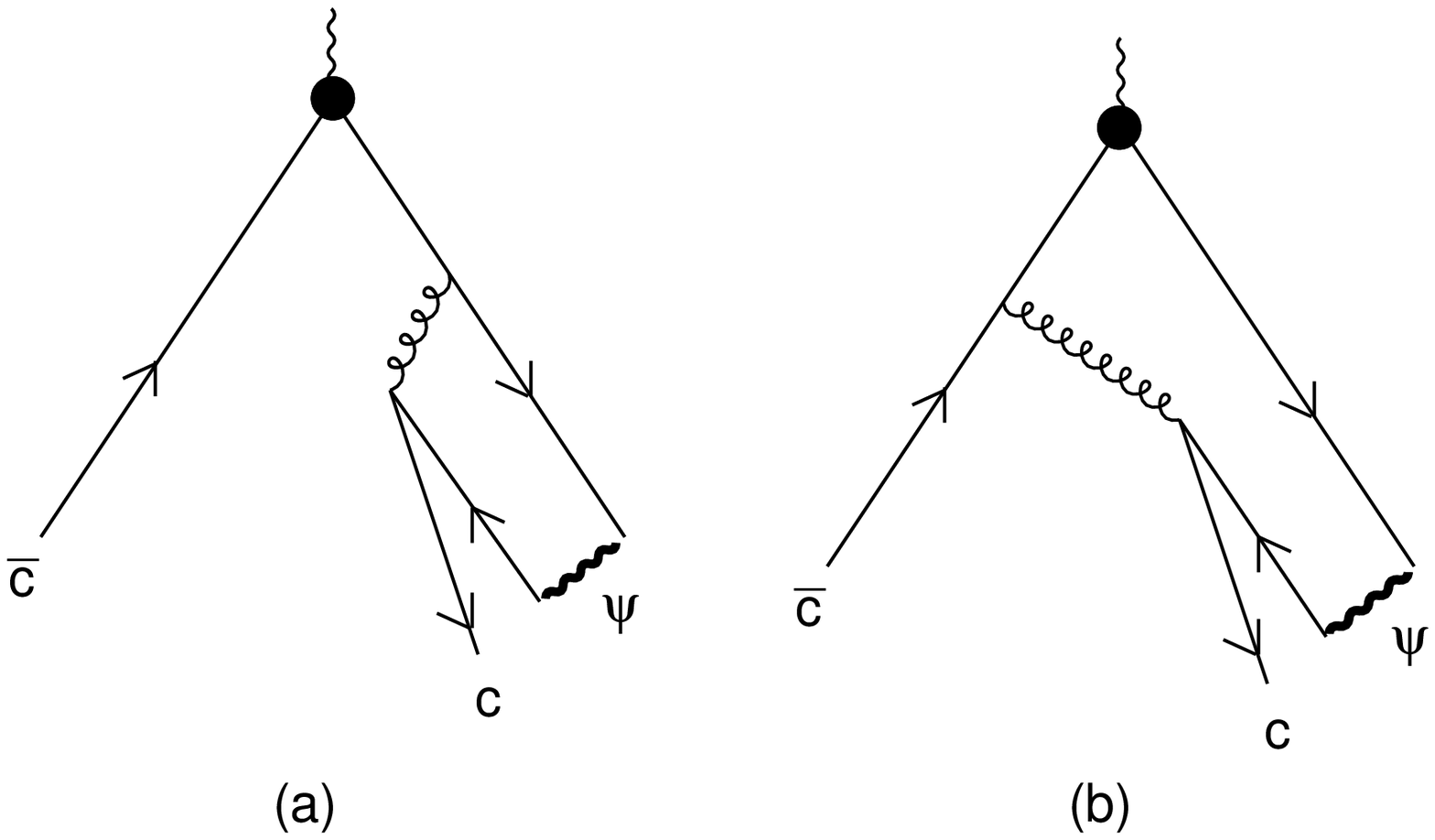}
\vfil\eject
\end